\documentclass[review]{elsarticle}
\usepackage{graphicx}
\usepackage{subfig}
\usepackage[noend]{algpseudocode}
\usepackage{amsmath,amssymb}
\usepackage{amsmath}
\usepackage{xcolor}
\usepackage{amsfonts}
\usepackage[ruled]{algorithm2e}
\usepackage{adjustbox}
\usepackage{multirow}
\usepackage{csquotes}
\usepackage{footmisc}
\usepackage{dblfloatfix}
\usepackage{float}
\newcommand\abs[1]{\left|#1\right|}

\journal{Journal of \LaTeX\ Templates}
\usepackage{nomencl}
\makenomenclature

\usepackage{etoolbox}
\renewcommand\nomgroup[1]{%
  \item[\bfseries
  \ifstrequal{#1}{A}{Abbreviations}{%
  \ifstrequal{#1}{S}{Sets}{%
  \ifstrequal{#1}{P}{Parameters}{
  \ifstrequal{#1}{V}{Variables}}}}%
]}

\begin{document}

\begin{frontmatter}

\title{Preference-based Energy Exchange in a Network of Microgrids}


\author[]{Li Bai\corref{mycorrespondingauthor}}
\cortext[mycorrespondingauthor]{Corresponding author}
\ead{baili123246530@gmail.com}

\author[]{Dimitri Thomopulos}
\author[]{Emanuele Crisostomi}

\address{Department of Energy, Systems, Territory and Constructions Engineering, \\University of Pisa, Pisa}

\begin{abstract}
Peer-to-peer energy trading is emerging as a new paradigm that in the near future may disrupt conventional electricity markets and heavily affect energy exchanges in networks of microgrids. In this paper, a preference mechanism is considered to compute optimal energy exchanges in a network of microgrids with or without the supervision of the distribution system operator, and the alternating direction method of multipliers is adopted for its distributed solution. The effect of the preference mechanism on the resulting power flow in the network is further studied and discussed for realistic case studies. Results show that a desired power flow in the network of interconnected microgrids can be achieved with different preference values locally chosen or imposed by the system operator. In particular, appropriate preferences may be used to give rise to different clusters of microgrids and reduce energy exchanges between different clusters.
\end{abstract}

\end{frontmatter}
\mbox{}
 
 

\nomenclature[A]{P2P}{Peer-to-peer}
\nomenclature[A]{MG}{Microgrid}
\nomenclature[A]{ADMM}{Alternating direction method of multipliers}
\nomenclature[A]{DSO}{Distribution system operator}
\nomenclature[A]{DER}{distributed energy resources}
\nomenclature[A]{RES}{Renewable energy resources}
\nomenclature[A]{ESS}{Energy storage systems}
\nomenclature[S]{$\Omega$}{Set collecting all MGs}
\nomenclature[S]{$\omega_n$}{Set collecting all trading MG partners of MG $n$}
\nomenclature[S]{$\mathcal{B}_n$}{Set collecting all buses in MG $n$}
\nomenclature[S]{$\mathcal{T}_n$}{Set collecting all distribution lines in MG $n$}
\nomenclature[S]{$\mathcal{G}_n$}{Set collecting all diesel generators (DGs) in MG $n$}
\nomenclature[S]{$\mathcal{S}_n$}{Set collecting all ESSs in MG $n$}
\nomenclature[S]{$\mathcal{E}$}{Set collecting all connecting edges within a netwok of microgrids}
\nomenclature[V]{$(i,j)$}{Distribution line connecting bus $i$ and $j$}
\nomenclature[V]{$p_{G,k}$}{Power generated by the $k$-th diesel generator $ k\in \mathcal{G}_n $}
\nomenclature[V]{$p_{\text{char},k}$}{Power charged by the $k$-th ESS, $k \in   \mathcal{S}_n$}
\nomenclature[V]{$p_{\text{disc},k}$}{Power discharged by the $k$-th ESS, $k \in   \mathcal{S}_n$}
\nomenclature[V]{$p_{\text{net},k}$}{Net load at the $k$-th bus $k \in \mathcal{B}_n$}
\nomenclature[V]{$P_i^{'}$}{Outlet active power at bus $i$}
\nomenclature[V]{$Q_i^{'}$}{Outlet reactive power at bus $i$}
\nomenclature[V]{$P_i$}{Active power of load at bus $i$}
\nomenclature[V]{$Q_i$}{Reactive power of load at bus $i$}
\nomenclature[V]{$V_i^{'}$}{Voltage at bus $i$}
\nomenclature[V]{$p_{nm}$}{Active power trade between MG $n$ and $m$}
\nomenclature[V]{$q_{nm}$}{Reactive power trade between MG $n$ and $m$}
\nomenclature[V]{$p_{n\text{DSO}}$}{Active power trade between MG $n$ and DSO}
\nomenclature[V]{$q_{n\text{DSO}}$}{Reactive power trade between MG $n$ and DSO}
\nomenclature[V]{$p_{nm}^{+}$}{Active power trade bought by MG $n$ from MG $m$}
\nomenclature[V]{$p_{nm}^{-}$}{Active power trade sold by MG $n$ to MG $m$}
\nomenclature[V]{$q_{nm}^{+}$}{Reactive power trade bought by MG $n$ from MG $m$}
\nomenclature[V]{$q_{nm}^{-}$}{Reactive power trade sold by MG $n$ to MG $m$}
\nomenclature[V]{$p_{n\text{DSO}}^{+}$}{Active power trade bought by MG $n$ from DSO}
\nomenclature[V]{$p_{n\text{DSO}}^{-}$}{Active power trade sold by MG $n$ to DSO}
\nomenclature[V]{$q_{n\text{DSO}}^{+}$}{Reactive power trade bought by MG $n$ from DSO}
\nomenclature[V]{$q_{n\text{DSO}}^{-}$}{Reactive power trade sold by MG $n$ to DSO}
\nomenclature[P]{$R_{ij}$}{Resistance of line $(i,j)$}
\nomenclature[P]{$X_{ij}$}{Reactance of line $(i,j)$}
\nomenclature[P]{$\lambda_{nm}$}{Preference of energy trade with MG $m$ for MG $n$}
\nomenclature[P]{$\lambda_{nn}$}{Local preference for MG $n$}
\nomenclature[P]{$\lambda_{n\text{DSO}}$}{Preference of energy trade with DSO for MG $n$}
\nomenclature[P]{$\kappa$}{Scaling parameter for preference terms}
\nomenclature[P]{$c_{\text{DSO}}$}{Electricity price for energy trade with DSO}
\nomenclature[P]{$c_{\text{loss}}$}{Electricity price for power loss on the distribution lines}
\nomenclature[P]{$T$}{Time horizon for P2P energy trading}
\nomenclature[P]{$N_{\Omega}$}{Size of set $\Omega$}
\printnomenclature
\section{Introduction}
\subsection{Motivation}
The increasing penetration of power generation from distributed energy resources (DERs), including wind and photo-voltaic (PV) plants, sometimes of the order of only a few kilowatts, are deeply affecting the conventional operation and architecture of traditional power systems. In this framework, the concept of microgrid (MG) is emerging as the basic unit at the foundation of a power grid \cite{Hamada2017}. Roughly speaking, a MG consists of several (possibly small-sized) DERs, energy storage systems (ESSs) (e.g., possibly including electric vehicles (EVs)), and controllable and uncontrollable loads. Most importantly, MGs have the ability to operate in grid-connected mode when connected to the external power grid, and also in islanding mode, if needed or more convenient. In this perspective, MGs can participate to electricity markets with the double-role of producers and consumers \cite{Paudel2018,ZIPP2017}.

At the same time, the concept of peer-to-peer (P2P) energy exchange is also emerging as a candidate alternative to current electricity markets, as it may be more suitable to fit a possible future scenario of a network of interconnected MGs. In a nutshell, P2P energy exchange corresponds to agents (e.g., here MGs) directly negotiating energy exchanges with each other, for instance through bilateral contracts, in the absence of third-party supervisors \cite{Moret2018,Baroche2019_1, ZHOU2018, TUSHAR2019}. As MGs may be interested in participating to the electricity markets only at occasional spots (i.e., when they do not operate in island mode), P2P energy transactions may be a convenient way to regulate such energy exchanges.

In principle, P2P energy exchanges, together with a MG-based architecture of a power grid, may revolutionize the conventional way in which power flows are currently computed. MGs may decide to exchange energy with other specific MGs in order to prioritize customized preferences (e.g., prioritizing neighbouring MGs to reduce the distance between power generation and consumption, or prioritizing energy generated from renewable sources), neglecting the actual topology of the underlying physical power grid. Accordingly, dynamic (i.e., time-varying) clusters of MGs exchanging energy among themselves may naturally arise at different moments in time. With this latter motivation in mind, as many works exist that investigate optimal energy exchanges in a network of MGs, but very few address the possibility of influencing the way optimal power flows are computed in practice, this paper shows how optimal power flows (optimal energy exchanges) may be computed (realized) to practically accomplish any specific topology of interest.
\subsection{State-of-art}
A network of MGs may either consist of grid-connected MGs, that mainly exchange energy with the main outer grid, or of interconnected MGs, that mainly exchange energy with each other \cite{Hans2019}. The first studies on the optimal energy management in networks of MGs initially investigated centralized approaches, where a central operator (sometimes called as an aggregator) gathered all available information required to compute optimal energy exchanges, as in \cite{Ouammi2015, Parisio2017}. Sometimes, e.g., \cite{Fathi2013}, it is envisaged that MGs should not merely execute the actions recommended by the aggregator, but should be intelligent enough to take autonomous actions and accomplish local objectives (e.g., minimize generation costs). This is expected to relieve the burden of the aggregator, in terms of computational and storage costs, which is one of the main drawbacks of centralized solutions.

As distributed algorithms are becoming very popular in the power systems community, as for instance the alternating direction method of multipliers (ADMM) or dual decomposition methods \cite{Molzahn2017}, hierarchical solutions \cite{Zhao2018} or decentralized solutions \cite{Gao2018, Feng2018, FANG2019} have been proposed as an alternative to purely centralized solutions. Here, only the relevant information is shared with the central node, and single MGs have local computational abilities to either work in the grid-connected mode or in an interconnected fashion \cite{Hans2019}. Also, both deterministic and stochastic optimization algorithms have been used to take into account the diverse models of DERs within a MG, as in \cite{Torres2019}. In hierarchical models, MGs can either buy or sell energy from the main grid using conventional retailer markets, or wholesale markets.

More recently, distributed peer-to-peer energy transactions are emerging as a new paradigm for energy exchange in the electricity markets, see for instance the very recent papers \cite{Moret2018,Torres2019,Sandgani2018,Jadhav2019, Sorin2019, Baroche2019}. In particular, different mechanisms have been proposed to include new facilities in P2P transactions. In \cite{Torres2019}, optimal P2P MG pairs are identified to optimize the cost function of each MG in the pair. In \cite{Jadhav2019}, historical power generation and load demand data are used to encourage MGs to trade with other MGs rather than with the main grid. In \cite{Sandgani2018}, a priority level is given to different clusters of MGs as the sequential order to solve a multi-objective problem. In \cite{Moret2018}, geographical preferences and the autonomy of communities are introduced to represent the willingness to exchange energy with the markets and with other community neighbors, respectively.  
In \cite{Sorin2019}, the framework of product differentiation is generally introduced in peer-to-peer trade and can be interpreted in terms of the bilateral trades. In \cite{Baroche2019}, differential grid cost allocation strategies are proposed in peer-to-peer energy transaction in terms of distance and zones. 

The intent of this paper is also in P2P energy exchanges in a network of MGs, but differently from the aforementioned papers, optimal energy exchange solutions are not obtained according to a specific utility function of interest. As power grids are complex physical systems, where different, and sometimes contrasting, aspects come into play, different optimal power flows can be obtained by tuning a few appropriate parameters, which is of our interest. In particular, by appropriately tuning the parameters of a preference matrix, it is possible to recover optimal power flows according to a P2P fully-connected power grid (which actually would not be convenient from the perspective of the power grid), grid-connected optimal power flows, or in principle any topology of optimal power flows of interest. Our proposed solution may be referred as a dynamic management of a network of MGs.

\subsection{Contribution}
In the framework of P2P energy transactions, a preference mechanism is embedded into the energy management model among interconnected MGs to achieve two main goals: first is to gain the ability to predetermine the topology of energy exchanges according to the preferences of the distribution system operator (DSO), second is to encourage more energy sharing among any pairs of MG traders than the pairs between MG and the main grid.

In this paper, the optimal energy management problem is solved in a network of MGs in a distributed way with an extra preference term. The presence of the extra term has two valuable benefits:
\newline
(1) It may be used by single MGs to decide from whom they would prefer to trade energy in a customized way (e.g., to prioritize energy generated by MGs based on renewable sources, or to buy energy from neighboring MGs);
\newline
(2) Preferences may be imposed by the DSO to choose the direction of energy flows and possibly create clusters of MGs that work in the islanding mode. It is known that this solution may be convenient for stability proposed from the perspective of DSO (\cite{Crisostomi2018, HOOSHMAND2019}).

In doing so, ADMM is applied to solve this power flow problem 
in a distributed way, which is a convenient strategy, as it allows MGs not to reveal possibly private information (e.g., amount of locally generated or consumed power). On the other hand, this poses challenging aspects in terms of the solution of mathematical optimal power flow problem, mainly due to the presence of tightly coupled equality constraints, as shown in greater details in section \ref{sec:admm}. Additionally, three cases studies are carried out to analyze the effect of such preference mechanism on the resulting power flow.



The remainder of the paper is organized as follows: section 2 describes the models of the distributed units included in a network of MGs and the objective; section 3 builds the overall preference-based optimization problem for a network of MGs; section 4 formulates the distributed method to solve the optimization problem; section 5 presents the results of three case studies; section 6 draws conclusions and outlines future works of our research.
\section{The Model of a MG}
A microgrid is typically made up of renewable energy sources (RESs), fuel-based energy sources, energy storage systems (ESSs) and loads \cite{MGbook2014}. In our MG model, RESs (e.g., PV and wind energy)  are considered as prioritized non-dispatchable units to maximize their usage and reduce environmental impact. Fuel generators (FGs) are dispatchable to ensure stability and offer more flexibility of operational modes to a MG. ESSs enhance reliability and resilience of a MG and contribute to balance supply and demand within the grid \cite{BANSHWAR2019}.
\subsection{FG Model}
A conventional model is considered for FGs in our MG. For every generic MG $n$, $\mathcal{G}_n$ denotes the set of all the FGs within it. The generation cost for $k$-th FG, $k\in \mathcal{G}_n$ for every generic MG $n$ is expressed as a quadratic function of the active output with the boundary constraints \cite{Zhao2018, Feng2018},
\begin{subequations}
	\begin{align}
		\label{eq:cons_fg_cost}
	&f_{G,k}(p_{G,k})=\frac{1}{2}a_k^2 p_{G,k}^2+b_k p_{G,k}+c_k, \\
	\label{eq:cons_fg_p}
	&\underline{p}_{G,k}\leq{p}_{G,k}\leq \overline{p}_{G,k}\\
	\label{eq:cons_fg_q}
	&\underline{q}_{G,k}\leq{q}_{G,k}\leq \overline{q}_{G,k}\\
	\label{eq:cons_fg_ramp}
	&-\gamma_{G,k}\leq{p}_{G,k}^{t+1}-{p}_{G,k}^{t}\leq \gamma_{G,k}
	\end{align}
\end{subequations}
where $p_{G,k}$ and $q_{G,k}$ are the generated active and reactive power,  nonnegative \(a_k\), \(b_k\) and \(c_k\) are the cost coefficients, and $\underline{p}_{G,k}$,  and $\overline{p}_{G,k}$  are the lower and upper bounds of the active power output. Similarly, $\underline{q}_{G,k}$,  and $\overline{q}_{G,k}$ are the counterparts for the reactive power. Finally, $\gamma_{G,k}$ represents the ramp rate bound. In our model, the time resolution is assumed to be one hour.
\subsection{ESS Model}
ESSs are modeled by a first-order discrete time model, accounting for the energy losses in the charging and discharging process \cite{XING2019, Crisostomi2019}. For every generic MG $n$, $\mathcal{S}_n$ denotes the set of all the ESSs within it. The model of $k$-th ESS, $k\in \mathcal{S}_n$ for every generic MG $n$ is expressed as
\begin{subequations}
\begin{align}
\label{eq:ess}
&x_k^{t+1}=x_k^{t}+\beta_{\text{char},k} p_{\text{char},k}^{t}-\beta_{\text{disc},k} p_{\text{disc},k}^{t}\\
\label{eq:ess_soc}
\quad &\underline{x}_k \leq  x_k^{t}\leq \overline{x}_k\\
\label{eq:ess_char_bds}
&0\leq p_{\text{char},k}^{t}\leq \overline{p}_{\text{char},k}\\
\label{eq:ess_disc_bds}
&0\leq p_{\text{disc},k}^{t}\leq \overline{p}_{\text{disc},k}
\end{align}
\end{subequations}
where \(x_k^{t}\) denotes the state of charge of ESS $k, k\in \mathcal{S}_n$ within MG \(n\) at time \(t\), \(\beta_{\text{char},k}\) and \(\beta_{\text{disc},k}\) denote the charging and discharging efficiency.
In addition, the state of charge \(x_k^{t}\), the charging power \(p_{\text{char},k}^{t}\) and discharging power \(p_{\text{disc},k}^{t}\) are bounded as shown in (\ref{eq:ess_soc})-(\ref{eq:ess_disc_bds}).
\subsection{RESs and Load Models}
MGs prioritize energy generated from RESs to fully satisfy energy demand. The loads and RES generation can be locally forecast ahead of time, taking account of local information such as numerical weather predictions and local consumer behaviors. A single variable of $p_{\text{net},k}$ is introduced as the net load, referring to the local loads offset by RESs generation connecting at the same bus $k$ in MG $n$.

\subsection{Distribution Line Model}\label{subsec:line}
As most of the distribution networks are radially designed, the \textit{DistFlow} model described in \cite{Baran1989} is a popular choice to model distribution lines. 
\begin{figure}[!htb]
    \centering
    \includegraphics[width=0.6\linewidth]{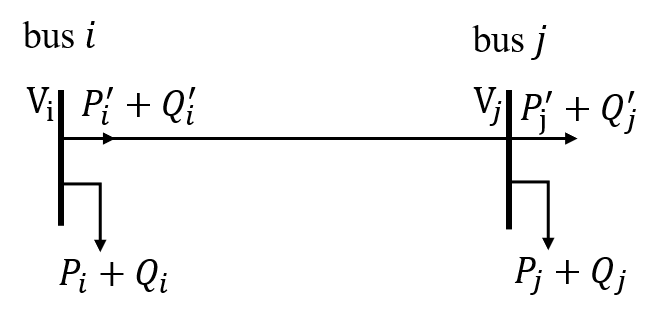}
    \caption{\textit{DistFlow} model}
    \label{fig:distflow}
\end{figure}
The \textit{DistFlow} model can be described for a distribution line $(i,j)$ connecting two neighboring buses $i$ and $j$, shown in Fig.\ref{fig:distflow}, as
\begin{subequations}
\begin{align}
    \label{eq:cons_flow_vol}
   &\mathit{v}_{i}=V_i^2, \quad \mathit{v}_{j}=V_j^2,\\
    \label{eq:cons_flow_cur}
    &P_{i}^{'2}+Q_{i}^{'2}=l_{ij}\mathit{v}_{i},\\
\label{eq:cons_flow_p}
    &P^{'}_{i}-P^{'}_{j}-R_{ij}l_{ij}-P_{j}=0,\\
    \label{eq:cons_flow_q}
    &Q^{'}_{i}-Q^{'}_{j}-X_{ij}l_{ij}-Q_{j}=0,\\
    \label{eq:cons_flow_v}
    &\mathit{v}_{i}-2(R_{ij}P_{i}^{'}+X_{ij}Q_{i}^{'})+(R_{ij}^2+X_{ij}^2)l_{ij}-\mathit{v}_{j}=0,\\
    \label{eq:cons_flow_loadpq}
    &P_{i}= P_{i}^{D}-P_{i}^{G}, \quad    Q_{j}= Q_{i}^{D}-Q_{i}^{G},  \quad P_{j}= P_{j}^{D}-P_{j}^{G}, \quad    Q_{j}= Q_{i}^{D}-Q_{i}^{G},\\
    \label{eq:cons_flow_bds}
    &\underline{P}^{'} \leq P^{'}_{i},P^{'}_{j}\leq \overline{P}^{'},\quad        \underline{Q}^{'} \leq Q^{'}_{i},Q^{'}_{j}\leq \overline{Q}^{'},\quad        \underline{V} \leq V_{i},V_{j}\leq \overline{V},
\end{align}
\end{subequations}
where $P^{'}_{i}$, $Q^{'}_{i}$ and $V_{i}$ represent the active outlet power, reactive outlet power and the voltage from the sending bus $i$, $P^{'}_{j}$, $Q^{'}_{j}$ and $V_{j}$ are the counterparts from the receiving end bus $j$, and $v_i$, $v_j$ and $l_{ij}$ are intermediary variables that derived from voltages and currents. All these variables are bounded as in (\ref{eq:cons_flow_bds}). In addition, $P_{j}+jQ_{j}$ represent the equivalent loads of the receiving bus $j$ by offsetting the local demand $P_{j}^{D}$ by local generation $P_{j}^{G}$. These two buses are connected by a branch line $(i, j)$ whose resistance and reactance are denoted by $R_{ij}$ and $X_{ij}$ respectively. 

To facilitate the solution of the optimal power flow problem, constraint (\ref{eq:cons_flow_cur}) is relaxed, as in \cite{Low2014}, into 
\begin{equation}
    \label{eq:relax}
    P_{i}^{'2}+Q_{i}^{'2} \leq \mathit{v}_i\mathit{l}_{ij}.
\end{equation}
This makes the optimization problem convex. As discussed in \cite{Low2014}, the optimal solution of optimal power flow problem is achieved when the relax constraint (\ref{eq:relax}) holds with equality. In addition, the power losses of distribution lines are also considered in the objective function.
\section{Energy Management of a Network of MGs}
Multiple individual MGs constitute a network of MGs by being physically connected within the distribution system. The physical connection provides the potential for P2P energy trades. With the aim of achieving P2P energy trades among a network of MGs and adjusting the power flow within the network, a preference mechanism is introduced in the optimization problem in the following. The overall objective function of a network of MGs is defined as the sum of the objectives of individual MGs. In the following description, though reactive power is not considered in the current day-ahead electricity market, it is still included in our model to provide further possibilities in trading reactive power in the future market in the distribution system. In the following, the set of all MGs in a nutshell will be denoted by $\Omega$, while the trading MG partners of MG $n$ will be denoted by $\omega_n$.

\subsection{Preference mechanism}
\label{subsec:pref}
A preference mechanism is introduced to optimize the energy exchange between interconnected MGs and customize the power flow in the physical network. Preferences can be decided by single MGs in a customized way or may be enforced by the DSO to achieve a predetermined configuration.

The preference on the energy trading for MG $n$ two MGs $n$ with MG $m, m \in \omega_n$ is given as $\lambda_{nm}$, and it adds a penalty term 
\begin{equation}
\label{eq:pref_terms}
    g(p_{nm})=\lambda_{nm}\abs{p_{nm}}
\end{equation}
to the objective function, where $\lambda_{nm}$ is nonnegative. In fact, the preference is adverse to its value, which indicates that a higher value contributes to a low preference. To better evaluate the different preference values for different traders of each MG, another parameter $\kappa$ and such constraints are introduced that
\begin{equation}
\begin{aligned}
    &\lambda_{nm}=\kappa\lambda_{nm}^{'},\\ & \sum_{m \in \omega_n}\lambda_{nm}^{'}=1, \\& \lambda_{nm}^{'}\geq 0,
\end{aligned}
\end{equation}
where $\lambda_{nm}^{'}$ and $\lambda_{nn}^{'}$ represent the normalized preference of power trade with MG $m$ and locally generated energy exchange energy for MG $n$, respectively. Parameter $\kappa$ is used to scale such normalized preferences into their corresponding trading prices $\lambda_{nm}$ and $\lambda_{nn}$. Parameter $\kappa$ may be either denoted as constant and fixed to mimic constant prices, or more realistically $\kappa$ may be time-varying to reflect typical electricity prices. Both simulations will be investigated later in section \ref{subsec:pref}. 
In the grid-connected mode, an extra preference on the energy trade with the main grid is introduced as $\lambda_{n\text{DSO}}=\kappa \lambda_{n\text{DSO}}^{'}$. Similarly, $\lambda_{n\text{DSO}}$ and $\lambda_{n\text{DSO}}^{'}$ are the actual and normalized preference terms respectively, where $\lambda_{n\text{DSO}}^{'}=1$ is used to prioritize energy trade between MG peers over the grid.
\subsection{Objective function}
The objective of a MG  consists of several terms including the generation costs of FGs, the cost of energy trade with the DSO and the preferences terms over a time horizon of $T$. In our paper, $T =24$ which is determined by day-ahead electricity market. The optimization problem for MG $n, n \in \Omega$ can be expressed as\\

\begin{subequations}
\label{eq:pro_one}
\begin{alignat}{3}
\label{eq:obj_mg_k}
&\!\min_{X_n}      &\qquad& L_n \\
&\text{s.t.} &     & L_n = \sum_{t=1}^{T_s}L_n^t \nonumber \\
&            &     & = \sum_{t=1}^{T_s}\big(c_{\text{DSO}}^{t}p_{n\text{DSO}}^{t}+ \sum_{k \in \mathcal{G}_n}f_{G,k}(p_{G,k}^{t})+c_{\text{loss}}^{t}\sum_{(i,j)\in \mathcal{T}_n}R_{ij}l_{ij}^t +g_{n}^{t}\big) \label{eq:obj_mg_k_sumt}\\
\label{eq:pref}
&            &     & g_n^{t}=\kappa\lambda_{n\text{DSO}}^{'}\abs{p_{n\text{DSO}}^{t}}+\kappa\lambda_{nn}^{'}\sum_{k \in \mathcal{S}_n}(p_{\text{char},k}^{t}+p_{\text{disc},k}^{t}) +\sum_{m\in \omega_n}
\kappa\lambda_{nm}^{'}\abs{p_{nm}^{t}}, \nonumber \\
&            &     & \forall t=1,\dots, T_s\\
\label{eq:bal}
&            &     & \sum_{k \in \mathcal{S}_n}\big(\beta_{\text{char},k}p_{\text{char},k}^{t}-\beta_{\text{disc},k}p_{\text{disc},k}^{t}\big)-\sum_{k\in \mathcal{G}_n}p_{G,k}^{t} \nonumber \\ 
&            &     & +\sum_{k \in \mathcal{B}_n}p_{\text{net},k}^{t}=\sum_{m\in \omega_n}p_{nm}^{t}+ p_{n\text{DSO}}^{t}, \quad \forall t=1,\dots, T_s \\
\label{eq:balq}
&            &     & -\sum_{k\in \mathcal{G}_n}q_{G,k}^{t}+\sum_{k \in \mathcal{B}_n}q_{\text{net},k}^{t}=\sum_{m\in \omega_n}q_{nm}^{t}+ q_{n\text{DSO}}^{t}, \quad \forall t=1,\dots, T_s \\
\label{eq:cons_local}
&            &     & (\ref{eq:cons_fg_cost})-(\ref{eq:cons_fg_ramp}), (\ref{eq:ess})-(\ref{eq:ess_disc_bds}), (\ref{eq:cons_flow_vol}),(\ref{eq:cons_flow_p})-(\ref{eq:cons_flow_bds}), (\ref{eq:relax}).
\end{alignat}
\end{subequations}

In the objective function (\ref{eq:obj_mg_k_sumt}), $c_{\text{DSO}}^{t}$ represents the unique electricity price regarding the power trade between DSO and any microgrid at time $t$. It is taken as the retail price in practice and can be forecast based on the historical price data. Besides, $c_{\text{loss}}^{t}$ represents the unique electricity price regarding the power loss on the distribution lines at time $t$. In principle, the price $c_{\text{loss}}^{t}$ regarding power loss on distribution lines, the price $c_{\text{DSO}}^{t}$ of exchange energy with the DSO and the scaling price $\kappa$ in the optimization problem may be different. Here, for the sake of simplicity, they are all considered to be equal and unique for any MG within a network.

Reviewing the preference terms in (\ref{eq:pref}), the objective function contains the absolute operator or $l_1$ norm regularization term. To simplify the objective function as a general quadratic function, extra variables are introduced to remove the absolute operator. As those variables are introduced for any time $t$, all notations by neglecting $t$ are simplified. Besides, the time step is generally one hour, and thus the amount of the power exchange is equal to the amount of the energy exchange. Specifically, $p_{nm}^{+}$ denotes the energy bought by MG $n$ from MG $m$, and  $p_{nm}^{-}$ denotes the energy sold by MG $n$ to MG $m$. Similarly, $p_{n\text{DSO}}^{+}$ and $p_{n\text{DSO}}^{-}$ denote the energy trades between MG $n$ and the main grid. Correspondingly,  $q_{nm}^{+}$,  $q_{nm}^{-}$, $q_{n\text{DSO}}^{+}$ and $q_{n\text{DSO}}^{-}$ are introduced as well. All those newly introduced variables are nonnegative, as
\begin{subequations}
\label{eq:nonneg}
    \begin{align}
    \label{eq:nonneg_p}
         &p_{nm}^{-}, \quad p_{nm}^{+}, \quad q_{nm}^{-},  \quad q_{nm}^{+} \geq 0\\
            \label{eq:nonneg_q}
         &p_{n\text{DSO}}^{-}, \quad p_{n\text{DSO}}^{+}, \quad
q_{n\text{DSO}}^{-},\quad q_{n\text{DSO}}^{+} \geq 0.
    \end{align}
\end{subequations}
Obviously,
\begin{subequations}
\label{eq:sum}
    \begin{align}
         &p_{nm}=p_{nm}^{+}-p_{nm}^{-},\\
         &q_{nm}=q_{nm}^{+}-q_{nm}^{-},\\
         &p_{n\text{DSO}}= p_{n\text{DSO}}^{+}-p_{n\text{DSO}}^{-},\\ &q_{n\text{DSO}}= q_{n\text{DSO}}^{+}-q_{n\text{DSO}}^{-}.
    \end{align}
\end{subequations}
The preference terms including absolute value operation can be rewritten, ignoring the superscript $t$, as
\begin{equation}
\label{eq:pref_noabs}
    g_n=\lambda_{n\text{DSO}}(p_{n\text{DSO}}^{+}+p_{n\text{DSO}}^{-})+\lambda_{nn}\sum_{k \in \mathcal{S}_n}(p_{\text{char},k}+p_{\text{disc},k})+\sum_{m\in \omega_n}
\lambda_{nm}(p_{nm}^{+}+p_{nm}^{-}).\\
\end{equation}
\textbf{Remark}: all newly introduced variables in (\ref{eq:nonneg}) and those regarding ESSs appear in the mutual preference terms and local preference terms in the objective function and in constraints as well. Since at a given time step, the energy exchange between 2 MGs can only have one direction, then one between $p_{nm}^{-}$ and $p_{nm}^{+}$ must be zero for MG $n$ (same for MG $m$). In other words, $p_{nm}^{-}p_{nm}^{+}=0$, $q_{nm}^{-}q_{nm}^{+}=0$, $p_{n\text{DSO}}^{-}p_{n\text{DSO}}^{+}=0$, $q_{n\text{DSO}}^{-}q_{n\text{DSO}}^{+}=0$. It can be proved that these constraints are automatically satisfied for the optimal solution. Besides, it can be explained intuitively that, if those constraints cannot be respected, then the objective value will be higher since all related variables contribute to the objective value in a positive manner (a similar discussion is also made in \cite{Hao2018}). Interestingly, it can be found that such preference terms naturally enforce a MG to be a seller or a buyer at one moment, rather than engage in arbitrage. Similarly, it also naturally enforce ESSs not work in the charging and discharging mode at the same moment.  Coincidentally, this is coherent with the assumptions proposed in \cite{Moret2018,Sorin2019, Crisostomi2019}.

In (\ref{eq:pro_one}), $X_n$ contains all local variables to be optimized, $L_n$ represents the local objective function, and all local constraints are further denoted, including (\ref{eq:bal})-(\ref{eq:cons_local}), (\ref{eq:nonneg}) and (\ref{eq:sum}), as the domain of definition $\text{Dom}_n$ for MG $n$. Therefore, a subproblem regarding MG $n$ in (\ref{eq:pro_one}) can be rewritten in a simple form as
\begin{equation}
\label{eq:pro_one_short}
    \min_{X_n} L_n,\quad X_n \in \text{Dom}_n.
\end{equation}

The optimization problem is constrained by the power balance of supply and demand of each MG, in addition to all the constraints determined by all the distributed units within a MG. As the DSO is not considered in the islanding mode, $p_{n\text{DSO}}^t$ is removed from the objective function and constraints as well. 

\section{Formulation of the Distributed Optimization Problem}\label{sec:admm}
In a centralized framework, an aggregator is required to determine the total power exchange with the main grid and the power allocations of all distributed units inside the distribution network, after gathering all cost curves of generators, forecasts of RESs and loads of all MGs, and the preferences of energy transactions as well. With the increasing penetration of DERs in the distribution network, each MG tends to be an autonomous entity for local energy control and management, and for energy transactions without sharing personal information such as the preferences of energy transactions, the local forecasts of load and RESs. More importantly, a distributed algorithm is highly required for achieving P2P energy trades among a network of MGs, as a P2P market is characterized by the lack of a supervisory agent \cite{Sorin2019}. 

The overall objective of a network of MGs is the sum of the individual objective of each MG. and it can be fully expressed as
\begin{equation}
L=\sum_{n \in \Omega}L_n.
\label{eq:OBJ_full}
\end{equation}
Additionally, $X=[X_0, \dots, X_{N_{\Omega}-1}]$ is the vector of all the variables to be optimized in the whole optimization problem, and $N_{\Omega}$ is the size of $\Omega$.

The overall objective built in (\ref{eq:OBJ_full}) can be accomplished in a distributed way by ADMM, as it can be separated for each MG with coupling constraints of the common physical connections and energy transactions. To solve the optimization problem in a distributed way, two kinds of coupling constraints must be respected: 1) the physical constraints are that the outlet power of the sending bus should equal the power of the receiving bus on the distribution line that physically connects two MGs; 2) the virtual constraints are that the energy transactions between any pair of MGs should be equal (i.e., the energy that MG $n$ buys from MG $m$ should be the same as  that MG $m$ sells to MG $n$). 

To implement ADMM, each agent (e.g., a MG) is required to share the energy of each peer trade and the physical variables (such as the active and reactive power flowing on the connecting line and the voltages of two buses on the two ends) regarding the connecting transmission line of a MG pair at each iteration of the algorithm. Obviously, the energy trade between any MG pairs of MG $n$ and $m$ should be equal, namely $p_{nm}=p_{mn}$, $q_{nm}=q_{mn}$. Let $\mathcal{E}=\{(n,m)\}$ denote the set of all edges connecting any MG pairs, where the edges physically refer to the distribution lines connecting any MG pairs. For example, edge $(n, m)$ indicates a distribution line that connects MG $n$ and MG $m$. As MGs $n$ and $m$ share the edge, they both make contributions to determine the power flows and voltages. Let us further introduce the shared physical variables for MG $n$ regarding the edge $(n, m)$: $\tilde{P}_{nm}^n$ represents the active power flowing on the edge at the sending end from MG $n$, $\tilde{Q}_{nm}^n$ represents the reactive power flowing on the edge at the sending end from MG $n$, $\tilde{V}_{n}^n$  and $\tilde{V}_{m}^n$ represent the voltages of the buses connected by the edge $(n, m)$. Accordingly, the shared variables of MG $m$ on the edge $(n, m)$ are introduced as $\tilde{P}_{nm}^{m}, \tilde{Q}_{nm}^{m}, \tilde{V}_{n}^{m}, \tilde{V}_{m}^{m}$. The shared variable vectors regarding each edge $(n,m)$ connecting MG $n$ and $m$, are further denoted as $Y_{n}=[\tilde{P}_{nm}^n, \tilde{Q}_{nm}^n, \tilde{V}_{n}^n, \tilde{V}_{m}^n]$ for MG $n$ and $Y_{m}=[\tilde{P}_{nm}^{m}, \tilde{Q}_{nm}^{m}, \tilde{V}_{n}^{m}, \tilde{V}_{m}^{m}]$ for MG $m$. Both $Y_n$ and $Y_m$ represent the physical variables for the same edge, therefore $Y_n=Y_m$.

Finally, the complete optimization problem can be rewritten as
\begin{equation}
\label{eq:vars_new}
\begin{aligned}
&\min_{X} & &\sum_{n \in \Omega}L_n,\\
&\text{s.t. } & &p_{nm} +p_{mn} =0 \quad\forall n, m \in \Omega,\\
&&&q_{nm} +q_{mn}=0 \quad\forall n, m\in \Omega,\\
&&&Y_{n} -Y_{m}=0 \quad\forall (n, m) \in \mathcal{E},\\
&&& X_n \in \text{Dom}_n.
 \end{aligned}
\end{equation}

The full optimization problem in (\ref{eq:vars_new}) is written in the classic form to be solved by ADMM algorithm in \cite{Boyd2011}.

\section{Simulation and Analysis}
\begin{figure}[!htb]
    \centering
    \includegraphics[width=0.9\linewidth]{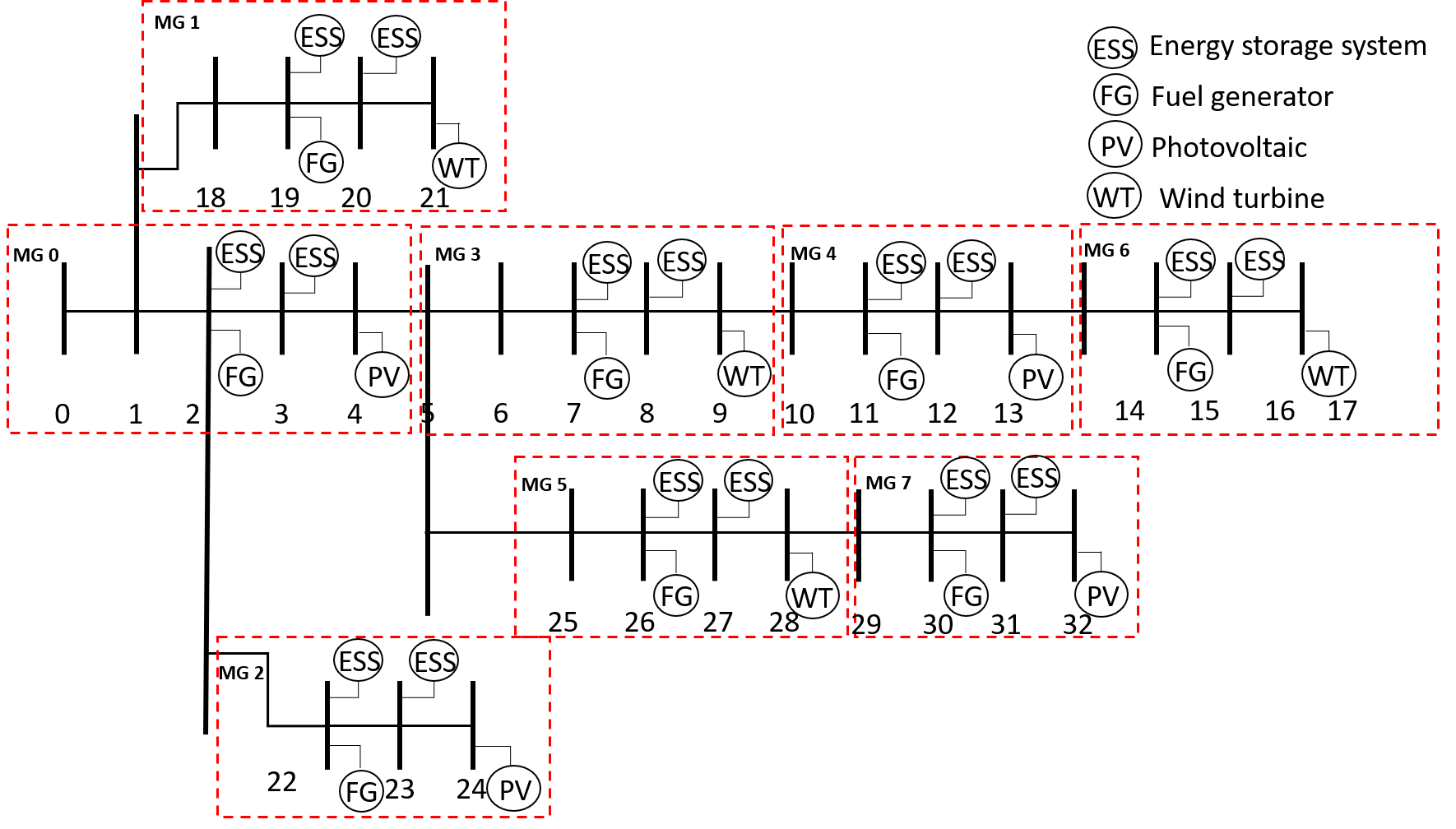}
    \caption{The topology of IEEE 33-node distribution system}
    \label{fig:33bus}
\end{figure}
\subsection{Description of the case study}
As for case study, the distributed optimal energy management is considered in the IEEE 33-node system \cite{HOOSHMAND2019,Wazir2016}. The radial network is decomposed into a network of 8 MGs, as in Fig.\ref{fig:33bus}. The network of MGs can be operated in the grid-connected mode by connecting bus 0 with the grid; otherwise it is in the islanding mode if bus 0 is disconnected from the grid.
\begin{figure}[!htb]
    \centering
    \includegraphics[width=0.8\linewidth]{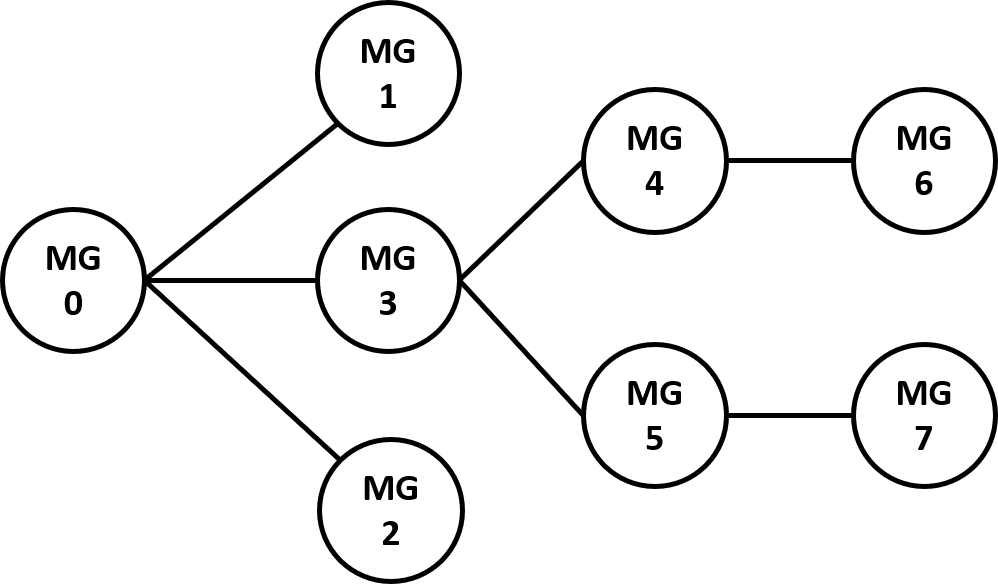}
    \caption{The equivalent graph of 8 MGs}
    \label{fig:33bus_8mgs}
\end{figure}

The physical topology of the network of MGs is presented in Fig.\ref{fig:33bus} and its equivalent schematic graph is displayed in Fig.\ref{fig:33bus_8mgs}. Two ESSs are considered in each MG to offer enough flexibility if there is excessive renewable generation. All electrical parameters regarding the transmission lines can be found in \cite{Wazir2016}, and the load and RES generations are modeled using the forecast data available\footnote{\label{note1}https://transparency.entsoe.eu/} to capture the daily features of the loads and RESs. In the grid-connected mode, the forecasts of retail price for the trade between MGs and DSO are made using the online electricity price data \cite{entsoe}.

With regard to the parameter configuration of a fuel generator, its capacity is 0.30MW and its ramping rate is 0.10 MW/h. Parameters $a_k$ and $b_k$ of FGs can be referred to the parameters of the small-capacity FGs in \cite{ZHAI2009}. The capacity of ESS is 0.2MW, and the charging and discharging efficiency are set to 0.9. The capacity of FGs are chosen large enough to usually cover all load consumption within a MG. 

\subsection{The effect of the preference matrix}\label{subsec:pref}
\begin{table*}[!htb]
    \caption{The preference matrices }
    \centering
    \begin{tabular}{c|cccccccccc}
        \hline
         \multirow{8}{*}{\parbox{1.2cm}{Case 0}}&&MG0 &MG1&MG2&MG3&MG4&MG5&MG6&MG7&grid\\
&MG0&0.02 &0.14 &0.14 &0.14 &0.14 &0.14 &0.14 &0.14& 1.0 \\
&MG1&0.14 &0.02 &0.14 &0.14 &0.14 &0.14 &0.14 &0.14& 1.0 \\
&MG2&0.14 &0.14 &0.02 &0.14 &0.14 &0.14 &0.14 &0.14& 1.0 \\
&MG3&0.14 &0.14 &0.14 &0.02 &0.14 &0.14 &0.14 &0.14& 1.0 \\
&MG4&0.14 &0.14 &0.14 &0.14 &0.02 &0.14 &0.14 &0.14& 1.0 \\
&MG5&0.14 &0.14 &0.14 &0.14 &0.14 &0.02 &0.14 &0.14& 1.0 \\
&MG6&0.14 &0.14 &0.14 &0.14 &0.14 &0.14 &0.02 &0.14& 1.0 \\
&MG7&0.14 &0.14 &0.14 &0.14 &0.14 &0.14 &0.14 &0.02& 1.0 \\
    \hline 
    \hline
         \multirow{8}{*}{\parbox{1.2cm}{Case 1}}&&MG0 &MG1&MG2&MG3&MG4&MG5&MG6&MG7&grid\\
&MG0&\textcolor{red}{0.02} &\textcolor{red}{0.04} &\textcolor{red}{0.04} &0.18 &0.18 &0.18 &0.18 &0.18& 1.0 \\
&MG1&\textcolor{red}{0.04} &\textcolor{red}{0.02} &\textcolor{red}{0.04} &0.18 &0.18 &0.18 &0.18 &0.18& 1.0 \\
&MG2&\textcolor{red}{0.04} &\textcolor{red}{0.04} &\textcolor{red}{0.02} &0.18 &0.18 &0.18 &0.18 &0.18& 1.0 \\
&MG3&0.30 &0.30 &0.30 &\textcolor{blue}{0.02} &\textcolor{blue}{0.02} &\textcolor{blue}{0.01} &\textcolor{blue}{0.03} &\textcolor{blue}{0.02} & 1.0\\
&MG4&0.30 &0.30 &0.30 &\textcolor{blue}{0.02} &\textcolor{blue}{0.01} &\textcolor{blue}{0.03} &\textcolor{blue}{0.02} &\textcolor{blue}{0.02}& 1.0 \\
&MG5&0.30 &0.30 &0.30 &\textcolor{blue}{0.01} &\textcolor{blue}{0.02} &\textcolor{blue}{0.02} &\textcolor{blue}{0.02} &\textcolor{blue}{0.03}& 1.0 \\
&MG6&0.30 &0.30 &0.30 &\textcolor{blue}{0.01} &\textcolor{blue}{0.02} &\textcolor{blue}{0.03} &\textcolor{blue}{0.02} &\textcolor{blue}{0.02}& 1.0 \\
&MG7&0.30 &0.30 &0.30 &\textcolor{blue}{0.02} &\textcolor{blue}{0.02} &\textcolor{blue}{0.02} &\textcolor{blue}{0.02} &\textcolor{blue}{0.02}& 1.0 \\
    \hline 
    \hline
\multirow{8}{*}{\parbox{1.2cm}{Case 2}}&&MG0 &MG1&MG2&MG3&MG4&MG5&MG6&MG7&grid\\
&MG0&\textcolor{red}{0.03} &\textcolor{red}{0.03} &\textcolor{red}{0.04} &0.18 &0.18 &0.18 &0.18 &0.18& 1.0\\
&MG1&\textcolor{red}{0.04} &\textcolor{red}{0.02} &\textcolor{red}{0.04} &0.18 &0.18 &0.18 &0.18 &0.18& 1.0 \\
&MG2&\textcolor{red}{0.03} &\textcolor{red}{0.05} &\textcolor{red}{0.02} &0.18 &0.18 &0.18 &0.18 &0.18& 1.0 \\
&MG3&0.18 &0.18 &0.18 &\textcolor{blue}{0.03} &\textcolor{blue}{0.03} &0.18 &\textcolor{blue}{0.04} &0.18&  1.0\\
&MG4&0.18 &0.18 &0.18 &\textcolor{blue}{0.05} &\textcolor{blue}{0.02} &0.18 &\textcolor{blue}{0.03} &0.18&  1.0\\
&MG5&0.15 &0.15 &0.15 &0.15 &0.15 &\textcolor{green}{0.03} &0.15 &\textcolor{green}{0.07}& 1.0 \\
&MG6&0.18 &0.18 &0.18 &\textcolor{blue}{0.04} &\textcolor{blue}{0.04} &0.18 &\textcolor{blue}{0.02} &0.18& 1.0 \\
&MG7&0.15 &0.15 &0.15 &0.15 &0.15 &\textcolor{green}{0.05} &0.15 &\textcolor{green}{0.05}& 1.0 \\
\hline
    \end{tabular}
    \label{tab:pref_mtx1}
    \end{table*}
One obvious benefit of the preference matrix is that it can be used to encapsulate information regarding preferences of single MGs, who may have their own preferences regarding whom they would like to buy (sell) power from (to), if needed. In addition, as seen in detail in this section, another advantage of the preference matrix, is that it may be imposed by the DSO to confine power flows within clusters of MGs, that end up working in island mode. This solution is known to be particularly convenient in terms of the stability of the power grids, and may be used when frequency oscillations reach critical values (in analogy to what had been done in \cite{Crisostomi2018}).

Given the same IEEE 33-node distribution system shown in Fig.\ref{fig:33bus}, the effect of three different preference matrices is considered, as presented in Table \ref{tab:pref_mtx1}. Reminding that a low value in the preference matrix corresponds to a high preference (and vice versa, as the preference matrix appears in the objective function to be minimized), it can be noticed that in the first case study equal preferences are considered. The second case study gives rise to two potential clusters, consisting of MGs 0-2 and 3-7 respectively. On the other hand, in the third case study, the DSO is interested in creating three independent islands of MGs, consisting of MGs 0-2; 3, 4, 6; and 5, 7 respectively. Such three case studies are examined, both when an outer power grid is considered (grid-connected mode), and also when it is not present (islanding mode). In the grid-connected mode $\lambda_{n\text{DSO}}=1$, which practically corresponds to prioritize power exchanges among MGs than with the power grid (to avoid having all MGs just exchanging power with the grid). 

First prices are assumed to be fixed, and the value of $\kappa$ is tuned at low, intermediate and high prices ($\kappa$ is equal to 20, 60 and 100 \$/MWh). Then a more realistic scenario of time-varying prices is considered, where $\kappa$ is randomly generated following a uniform distribution in the interval of $[20, 100]$\$/MWh. In general, when the price is low, MGs may prioritize power generated at a lower price by other MGs (e.g., from renewable sources) over more expensive power locally generated (e.g., from conventional generators). On the other hand, when the price is high, it may be more convenient to locally generated the required power.

\begin{figure}[!htb]
    \centering
    \includegraphics[width=0.8\linewidth]{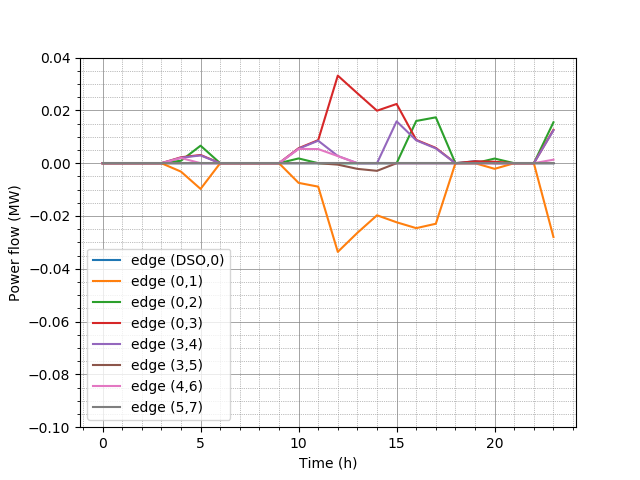}
    \caption{ Power flows on the edges in the grid-connected mode in Case 1 }
    \label{fig:case1_dso}
\end{figure}
\begin{figure*}[!htb]
    \centering
    \subfloat[$\kappa=20$]{\label{fig:case1_20}{\includegraphics[width=0.45\linewidth]{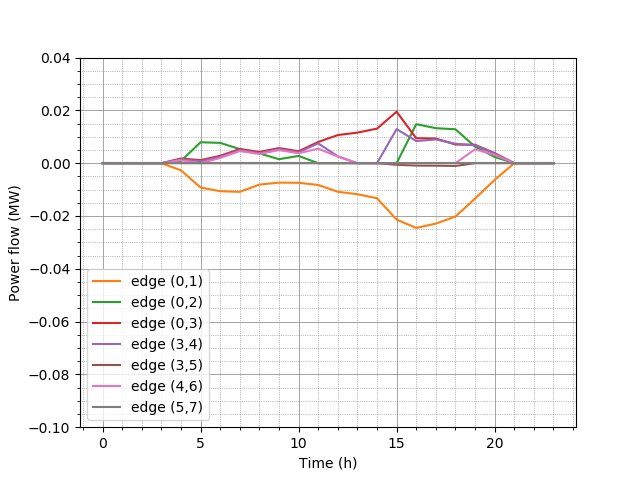} }}%
    \quad
    \subfloat[$\kappa=60$]{\label{fig:case1_60}{\includegraphics[width=0.45\linewidth]{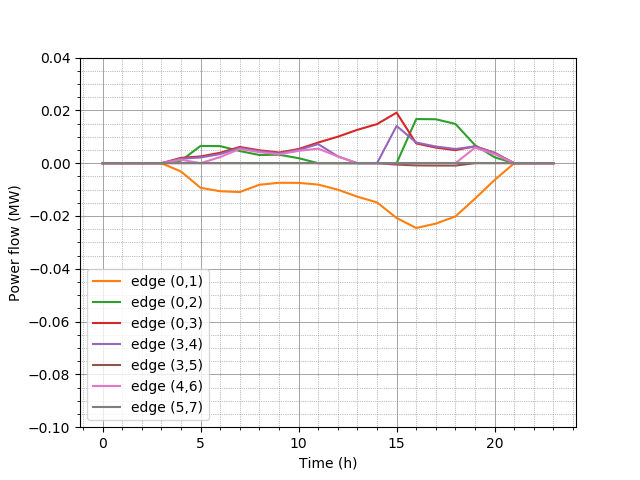} }}%
    \\
        \subfloat[$\kappa=100$]{ \label{fig:case1_100}{\includegraphics[width=0.45\linewidth]{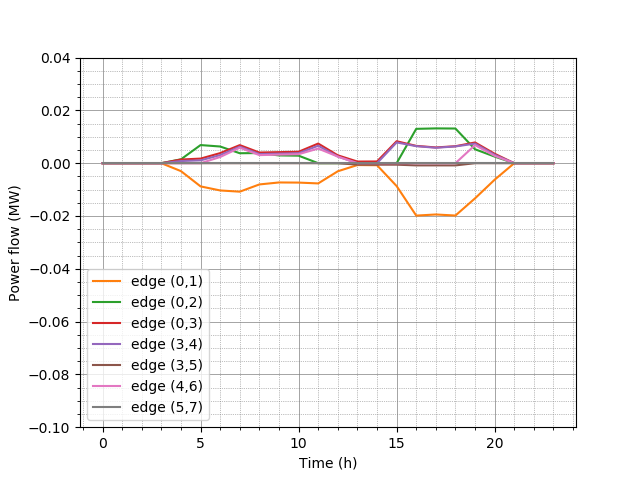} }}%
    \quad
    \subfloat[a time-varying $\kappa$ ]{ \label{fig:case1_random}{\includegraphics[width=0.45\linewidth]{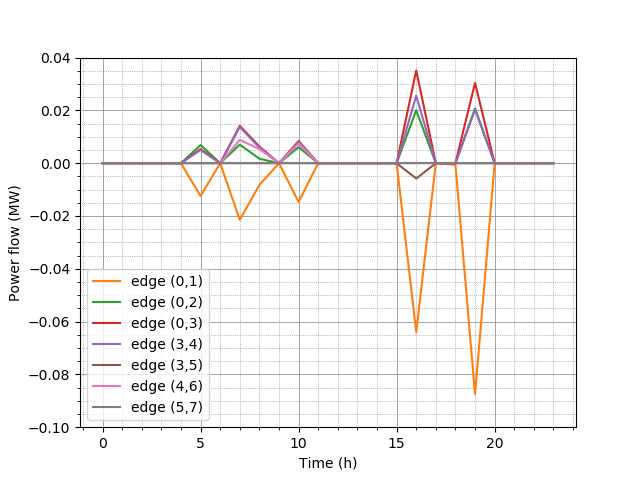} }}%
    \caption{Power flows on the edges in the islanding mode in case 1}%
    \label{fig:case1}%
\end{figure*}
\subsection{Case studies}
Three case studies are carried out according to the three preference matrices shown in Table \ref{tab:pref_mtx1}. In all the cases, the values of the objective functions obtained using the ADMM algorithm are almost the same to those obtained using a centralized algorithm. It indicates that the ADMM algorithm successfully converges to the optimal solution.

In three different case studies, the first one can be a reference point for the latter two case studies, which helps to illustrate the function of the preference matrix on power flow and power trading. In addition, among the legends displayed in the following figures, \enquote{edge (DSO,0)} represents the edge connecting MG 0 with DSO, and similarly \enquote{edge (n,m)} represents the edge connecting MG $n$ with MG $m$. 

\subsubsection{Case 1: equal preferences} 
In the first case study, equal preferences among all MGs are considered. As previously mentioned, self-consumption is prioritized and power exchanges with the DSO are penalized. This case study basically corresponds to the typical optimal problem of energy exchanges in a network of MGs that has been already widely investigated in the literature. 
In the grid-connected mode, all power flows on the edges are displayed in Fig.\ref{fig:case1_dso}. It can be observed that such a network of microgrid operates in the islanding mode despite physical connection to the DSO, as power flowing on the edge (DSO, 0) is zero for the whole time horizon (i.e., 24 hours). This is a consequence of penalizing energy exchange with the DSO. In addition, the power flowing on the edge (5,7) is also zero, indicating that MG 7 is self-sufficient and operates in islanding mode. Also, in the islanding mode, the results presented in Fig.\ref{fig:case1} show that MG 7 does not exchange power with the other MGs. Given the same preferences,  Figs.\ref{fig:case1_20}-\ref{fig:case1_100} show that power flows, although having similar trends, decrease with the increase of parameter $\kappa$. A time-varying price gives rise to completely different power flows, as shown in Fig.\ref{fig:case1_random}. In particular, it can be observed that energy exchanges largely depend on time-varying electricity prices (randomly generated in the interval of [20,100] \$/MWh). For instance, a fairly low price was considered at 19:00 and all MGs take advantages of this convenient price to exchange energy at that time of that day. 

\begin{figure}[!htb]
    \centering
    \includegraphics[width=0.8\linewidth]{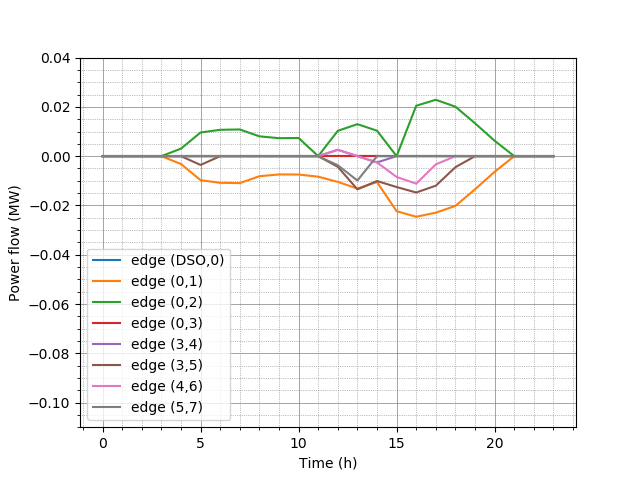}
    \caption{ Power flows on the edges in the grid-connected mode in Case 2 }
    \label{fig:case2_dso}
\end{figure}

\begin{figure*}[!htb]
    \centering
    \subfloat[$\kappa=20$]{\label{fig:case2_20}{\includegraphics[width=0.45\linewidth]{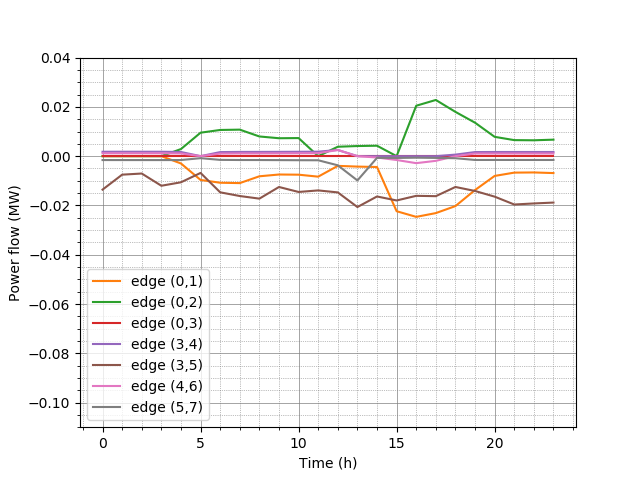} }}%
    \quad
    \subfloat[$\kappa=60$]{\label{fig:case2_60}{\includegraphics[width=0.45\linewidth]{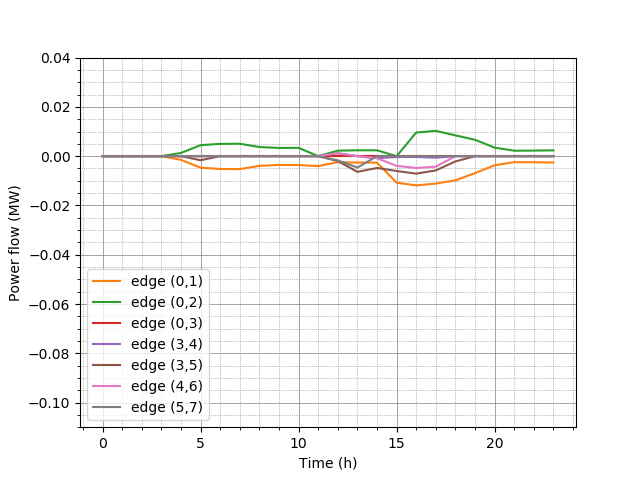} }}%
    \\
        \subfloat[$\kappa=100$]{ \label{fig:case2_100}{\includegraphics[width=0.45\linewidth]{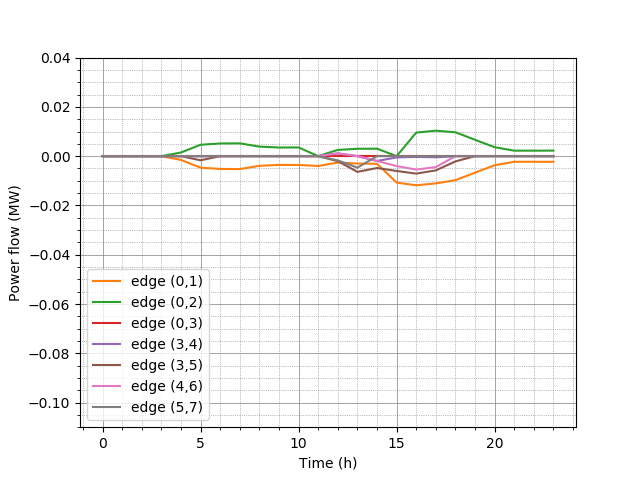} }}%
    \quad
    \subfloat[a time-varying $\kappa$]{ \label{fig:case2_random}{\includegraphics[width=0.45\linewidth]{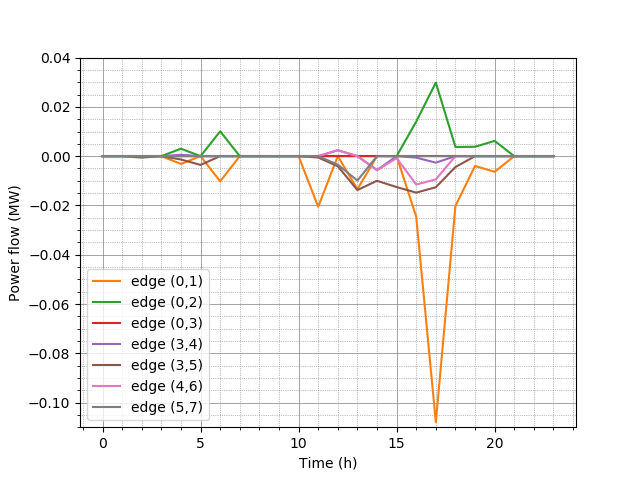} }}%
    \caption{Power flows on the edges in the islanding mode in case 2}%
    \label{fig:case2}%
\end{figure*}

\subsubsection{Case 2: 2 clusters} 
In this case study, the preference matrix aims at creating two clusters of MGs, namely, one including MGs 0-2, and the one including MGs 3-7. The results of the grid-connected case are displayed in Fig.\ref{fig:case2_dso}. It can be immediately noticed that the power flow along the edge (0,3) is zero for the whole horizon of 24 hours, which practically confirms that this preference matrix managed to divide the network of MGs into two clusters that do not exchange power among themselves (as edge (0,3) is the only link that connects such two clusters). In addition, since $\lambda_{n\text{DSO}}=1$, it can be seen that also the power along the transmission line between the power grid and bus 0 is almost 0 most of the time. Slightly differently from case 1, power flows on the edge (5,7) is not zero as lower preference values are provided to the trades involving MG 7.

Now the case is investigated when the outer power grid is not considered, and the 8 MGs are only interconnected among themselves. In this case,  Figs \ref{fig:case2_20}-\ref{fig:case2_100} show power flows when $\kappa$ is constant and equal to 20, 60 and 100 \$/MWh respectively. It can be observed that as before fewer energy exchanges occur when prices are higher.
Finally, Fig.\ref{fig:case2_random} shows the results regarding a time-varying electricity price (randomly generated in the interval of $[20,100]$ \$/MWh). While energy exchanges are completely different, reflecting the time-varying prices (e.g. many energy exchanges occur at 17:00 when random prices are low), it can be still observed that two clusters are preserved as expected.
 \begin{figure}[!htb]
    \centering
    \includegraphics[width=0.8\linewidth]{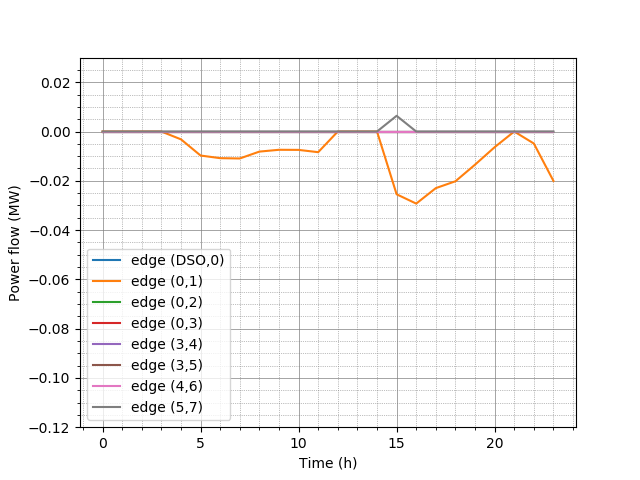}
    \caption{Power flows on the edges in the grid-connected mode in Case 3}
    \label{fig:case3_dso}
\end{figure} 
\begin{figure*}[!htb]
    \centering
    \subfloat[$\kappa=20$]{\label{fig:case3_20}{\includegraphics[width=0.45\linewidth]{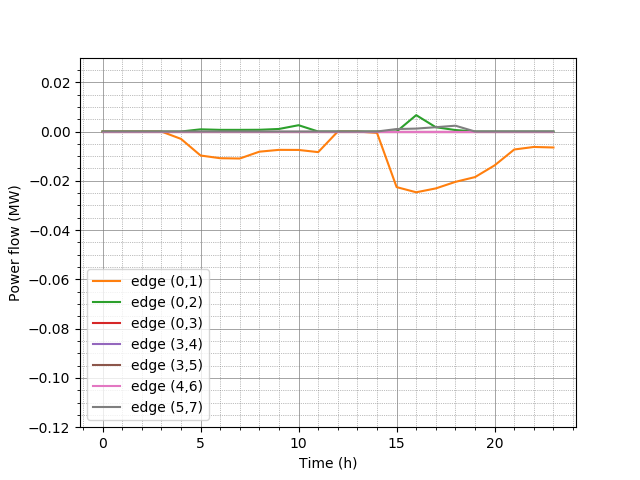} }}%
    \quad
    \subfloat[$\kappa=60$]{\label{fig:case3_60}{\includegraphics[width=0.45\linewidth]{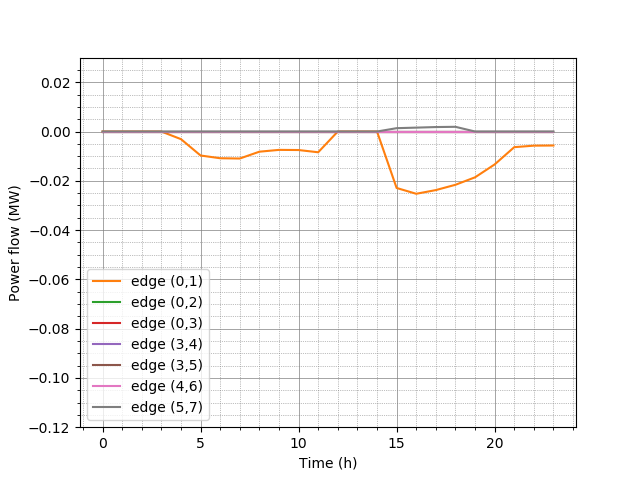} }}%
    \\
   \subfloat[$\kappa=100$]{\label{fig:case3_100}{\includegraphics[width=0.45\linewidth]{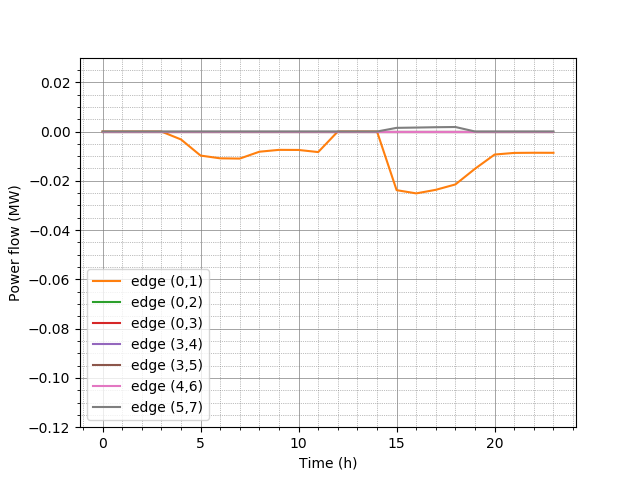} }}%
    \quad
    \subfloat[a time-varying $\kappa$]{\label{fig:case3_random}{\includegraphics[width=0.45\linewidth]{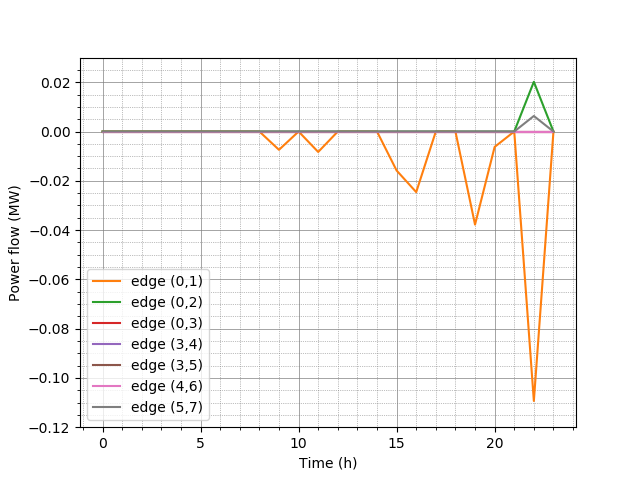} }}%
    \caption{Power flows on the edges in the islanding mode in case 3}%
    \label{fig:case3}%
\end{figure*}
\subsubsection{Case 3: 3 clusters} The third preference matrix of Table \ref{tab:pref_mtx1} is considered, which may give rise to 3 different clusters, and recompute the optimal power flows under the same assumptions of the previous case studies. In the grid-connected case shown in Fig.\ref{fig:case3_dso}, it can be observed that only power flows along the edges (0,1) and (5,7) within the time horizon. This implies that most of the microgrids within the network work in the islanding mode, including MGs 2, 3, 4 and 6, while MGs 0 and 1 form a cluster and MGs 5 and 7 form another one. Obviously, the 3 expected clusters are achieved.

Similarly to the previous case studies, Figs.\ref{fig:case3_20}-\ref{fig:case3_random} show what happens when only MGs are considered in the network, for 3 different constant prices, and in the more realistic time-varying electricity price. In particular, the 3 clusters are well established under such different pricing scenarios. 

While many other case studies may be artificially devised by appropriately tuning the entries of the preference matrix, the latter two case studies are to emphasize the influence of the preference matrix in shaping optimal power flows as desired. Finally, it can be observed again that fewer energy exchanges occur when higher prices are considered, and that energy exchanges are more irregular and and concentrated at favorable  (when the price is lower) in the last scenario with varying price. 

\section{Conclusion}
In this paper, a novel optimization problem with a preference mechanism was formulated to optimally solve the energy management problem in a network of microgrids. In particular, the alternating direction method of multipliers was used to solve the problem in a distributed fashion, thus allowing microgrids not to reveal some possibly private and sensitive information, while still respecting all the constraints of the underlying shared distribution network. In particular, we showed that the preference mechanism may be used by the distribution network operator to shape the topology with which power flows in the power grid, and in particular to conveniently give rise to clusters of microgrids \cite{Crisostomi2018}. Finally, the solution was shown to be efficient both for grid-connected infrastructures, and also when the power network is only formed by interconnected microgrids.

As very few other works investigate the impact of peer-to-peer energy exchanges in networks of microgrids down to the level of power flows, this paper may be seen as a first step in that direction. But as the notion of microgrids is continuously evolving, and discussions about future electricity markets including peer-to-peer aspects are gaining momentum, there appears to be the demand of new mathematical tools and new strategies to conveniently formulate and solve energy exchange problems. 
\section{Acknowledgment}
The authors would like to thank Pierre Pinson and Fabio Moret (Technical University of Denmark) for sharing their valuable thoughts and ideas on this topic.

\section{References}
\bibliography{mybibfile}
\end{document}